\documentclass[preprint,12pt]{elsarticle}

\usepackage{amssymb}
\usepackage{amsmath}

\journal{Journal of Subatomic Particles and Cosmology}

\begin{document}

\begin{frontmatter}



\title{$\rm K_{1}(1270)$ production in pp collisions with ALICE} 


\author{Su-Jeong Ji for the ALICE Collaboration} 

\affiliation{organization={Department of Physics, Pusan National University},
            city={Busan},
            postcode={46241}, 
            country={South Korea}}

\begin{abstract}
The study of chiral partners such as $\rm{K_1}$ and $\rm{K^*}$ mesons, which have vacuum widths below 100 MeV, provides an opportunity to explore chiral symmetry restoration in heavy-ion collisions. A recent theoretical model suggests that the $\rm{K_1/K^*}$ ratio in heavy-ion collisions should be significantly larger than that predicted by the statistical hadronisation model. Investigating the $\rm{K_1/K^*}$ ratio as a function of multiplicity across various collision systems, from pp to central heavy-ion collisions, can shed light on the impact of chiral symmetry restoration. However, the $\rm{K_1}$ meson has yet to be observed in hadron-hadron collisions.

The ALICE detector’s advanced particle identification capabilities enable the measurement of the $\rm{K_1}$ meson through hadronic decay modes, such as  $\rm{K_{1}^{-} \rightarrow \rho^{0}K^{-}}$  and $\rm{K_{1}^{-} \rightarrow \pi^{-}\bar{K}^{*0}}$. This presentation discusses the feasibility of measuring $\rm{K_1}$ mesons in pp collisions using the ALICE experiment.
\end{abstract}




\end{frontmatter}



\section{Introduction}
\label{sec1}


Relativistic heavy-ion collisions create a distinctive environment for exploring nuclear matter subjected to extreme temperatures and densities. At sufficiently high energy density ($\sim$ 1 GeV/$\mathrm{fm^3}$), a quark-gluon plasma (QGP) is expected to be created, where quarks and gluons are no longer confined within hadrons. It is also expected that the spontaneous breaking of chiral symmetry can be restored under these extreme conditions. As the system undergoes rapid expansion and cooling, the QGP goes back into a hadronic gas, where the hadrons are formed and interact with each other. In this stage, short-lived hadrons whose lifetimes are around a few fm/$c$ can undergo rescattering and regeneration processes, making us unable to know the exact number of produced resonances. 
Therefore, comparing the number of produced short-lived resonances and stable hadrons or their elliptic flow are important for studying the lifetime of the hadronic phase. As $\rm{K_1}(1270)$ has a lifetime of around 2 fm/$c$, which is between the well-known particles, $\rm{\rho^0}$ and $\rm{K^*}$, studying of $\rm{K_1}$ can be important for confining the lifetime of hadronic phase. 

\begin{figure}[htb]
\centering
\includegraphics[width=1.0\textwidth]{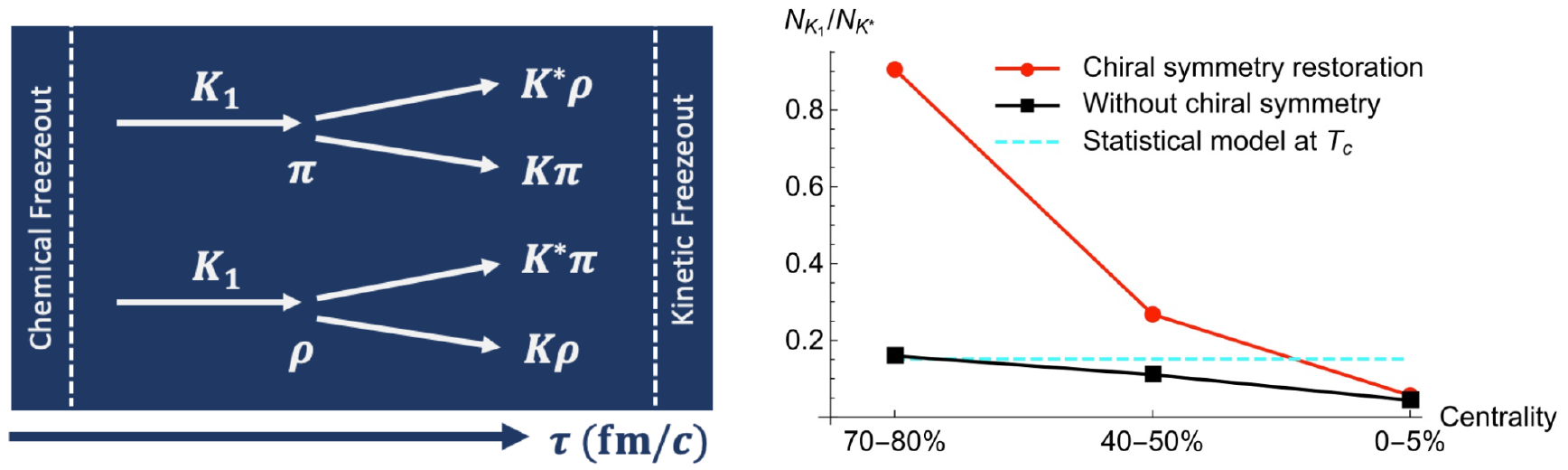}
\caption{Possible inelastic interaction in hadronic phase of $\rm{K_1}$ (left) and comparison of statistical hadronisation model and theoretical calculation of $N_{\rm{K_1}}/N_{\rm{K^*}}$ in heavy-ion collisions with and without chiral symmetry restoration effects as a function of centrality~\citep{sung2021k1}.}\label{fig1}
\end{figure}

It is expected that the spontaneous breaking of chiral symmetry can be restored in the QGP. $\rm{K_1}(1270)$ and $\rm{K^*(892)}$, whose vacuum widths are $90\pm20$ MeV and $47.3\pm0.5$ MeV are one of the strong candidates to study the chiral symmetry restoration with their narrow widths~\citep{ParticleDataGroup:2024cfk}. A recent theoretical calculation indicates that the $\rm{K_1/K^*}$ ratio in heavy-ion collisions is expected to be considerably higher than the values predicted by the statistical hadronisation model, as shown in Fig.~\ref{fig1} (right). However, due to the large number of produced hadrons in central heavy-ion collisions, both $\rm{K_1}$ and $\rm{K^*}$ particles will undergo inelastic scattering with light hadrons during hardonic phase, ending up suppression of the yield of $\rm{K_1}$ and relatively enhancing the yield of $\rm{K^*}$. Therefore, measuring $\rm{K_1/K^*}$ ratio as a function of multiplicity across various collision systems, valuable insights into the effects of chiral symmetry restoration and rescattering in hardonic phase can be obtained. However, as $\rm{K_1}(1270)$ has not been measured in hadron-hadron collisions yet, we initiated our research by measuring $\rm{K_1}$ in pp collisions and then will expand the investigation to different collision systems. This contribution presents first look of the measurement of $\rm{K_1}(1270)$ using the decay channels $\rm{K_{1}^{-} \rightarrow \rho^{0}K^{-}}$  and $\rm{K_{1}^{-} \rightarrow \pi^{-}\bar{K}^{*0}}$ in pp collisions in 50--100\% low multiplicity events with ALICE detectors~\citep{alice2014performance}.

\section{Feasibility study in ALICE}
\label{sec2}

The ALICE detector at Large Hadron Collider (LHC) at CERN is designed to study heavy-ion collisions.
In this analysis, main subdetectors such as Time Projection Chamber (TPC) and the Time-of-Flight (TOF) detectors are used, which enable precise tracking and particle identification. 
The TPC provides high resolution tracking performance and particle identification via energy loss, while TOF helps identifying via measuring velocity of charged particles. 
Additionally, the Inner Tracking System (ITS) plays a main role as vertex reconstruction, improving the identification of secondary decays. 
These features of ALICE detectors allow to reconstruct $\rm{K_1}$ meson through their hadron decay modes, making it possible to investigate $\rm{K_1}$ production in pp collisions.

\begin{figure}[htb]
\centering
\includegraphics[width=1.0\textwidth]{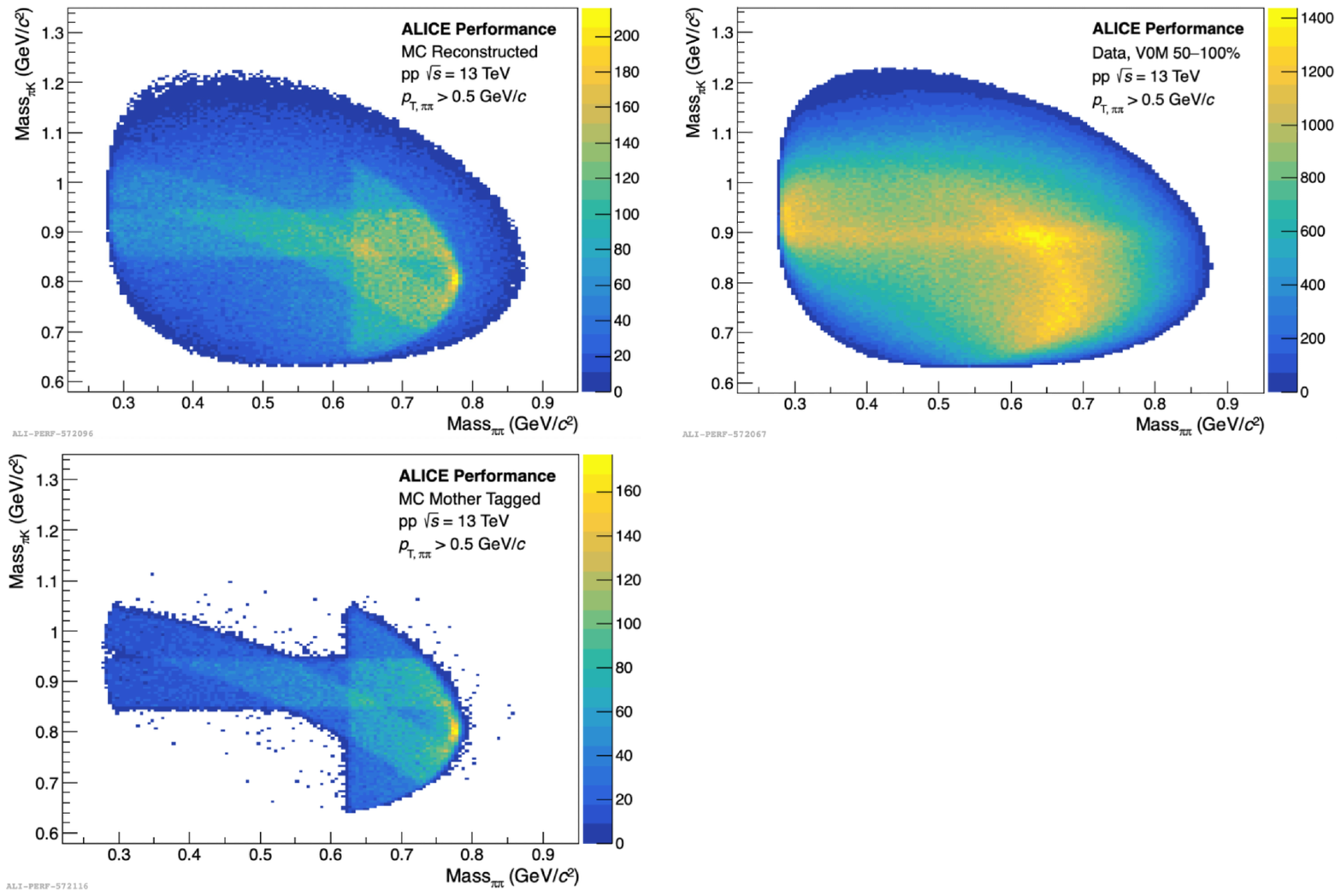}
\caption{2-dimensional invariant mass distribution of $\rm{\pi} \rm{K}$ versus $\rm{\pi} \rm{\pi}$ from Monte-Carlo reconstructed level (left top), mother tagged (left bottom), and ALICE data (right) within the mass range of $1.170 < M_{\rm{\pi\pi K}}<1.370$ GeV/$c^2$.}\label{fig2}
\end{figure}

The $\rm{K_1}$ is reconstructed using the decay channels $\rm{K_{1}^{-} \rightarrow \rho^{0}K^{-}}$ and
$\rm{K}_{1}^{-} \rightarrow $ $\pi^{-}\rm{\bar{K}}^{*0}$, resulting in $\pi\pi\rm{K}$ combinations.
Prior to analysing the invariant mass distribution of $\pi\pi\rm{K}$ pairs, we examine the two-dimensional invariant mass distribution of $\rm{\pi K}$ versus $\rm{\pi\pi}$ to evaluate the abundance of $\rm{K_1}$ candidates within the mass range of $1.170 < M_{\rm{\pi\pi K}} < 1.370$ GeV/$c^2$, as shown in Fig.~\ref{fig2}.

\begin{figure}[htb]
\centering
\includegraphics[width=1.0\textwidth]{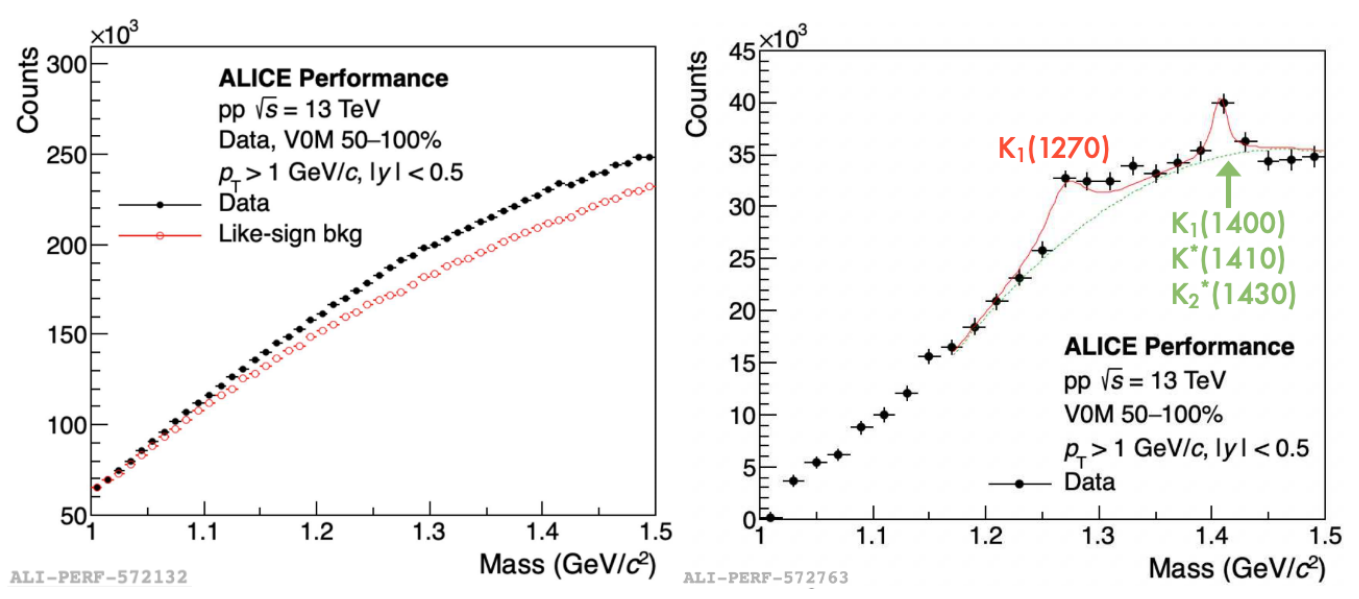}
\caption{The invariant mass distribution of $\pi\pi\rm{K}$ pairs' unlike-sign in black and like-sign background in red (left) and after subtraction like-sign from unlike-sign (right).}\label{fig3}
\end{figure}

Fig.~\ref{fig2} compares the Monte Carlo reconstructed level with the ALICE data. The Monte Carlo events are simulated using the PYTHIA8~\citep{PYTHIA8} event generator with the injection of $\rm{K_1}$ signals. The structures observed in the Monte Carlo reconstruction are directly reflected in the mother-tagged plot and exhibit similar patterns to those in the experimental data. This alignment confirms the significant production of $\rm{K_1}$. 
We proceeded to analyse the invariant mass distribution of $\pi\pi\rm{K}$ pairs, as illustrated in Fig.~\ref{fig3}. The left panel presents unlike-sign pairs ($\rm{\pi^\pm\pi^\mp K^\pm}$) marked with black markers and like-sign pairs ($\rm{\pi^\pm\pi^\pm K^\mp}$ and $\rm{\pi^\pm\pi^\pm K^\pm}$) marked with red markers. By subtracting the like-sign distribution from the unlike-sign distribution, the resulting signal is shown at right panel which exhibits peak structures near 1270 MeV/$c^2$, consistent with the $\rm{K_1}$, and additional structures around 1400 MeV/$c^2$, which can be attributed to the combination of $\rm{K_1}(1400)$, $\rm{K^*}(1410)$ and $\rm{K^*_{2}}(1430)$. The invariant mass spectrum was fitted using two Breit-Wigner functions to model the signal peaks and a quadratic function to account for the background.

These results provide clear evidence for the presence of $\rm{K_1}$ resonances within the analysed multiplicity range, validating our approach to reconstruct $\rm{K_1}$ in pp collisions. This result paves the way for extending the study to different multiplicity regions with different collision systems, enhancing our understanding of $\rm{K_1}$ production and enabling further studies.

\section{Summary}

In summary, we present the initial measurements of $\rm{K_1}(1270)$ decays into $\rm{K_{1}^{-} \rightarrow \rho^{0}K^{-}}$ and $\rm{K_{1}^{-} \rightarrow \pi^{-}\bar{K}^{*0}}$ in pp collisions at low multiplicity (50–100\%) using the ALICE detector. The $\rm{K_1}$ was reconstructed through $\pi\pi\rm{K}$ combinations, and the invariant mass distributions were analysed. The comparison between Monte Carlo simulations and ALICE data confirmed the significant production of $\rm{K_1}$ resonances. The invariant mass spectra exhibited clear peak structures around 1270 MeV/$c^2$ and 1400 MeV/$c^2$, consistent with theoretical expectations. This study validates our methodology for reconstructing $\rm{K_1}$ in pp collisions and lay the groundwork for expanding the study to different multiplicity regions and collision systems, thereby enhancing our understanding of $\rm{K_1}$ production mechanisms and the dynamics of the hadronic phase.

\clearpage

\section*{Acknowledgment}
The work was supported by the National Research Foundation of Korea (NRF) grant funded by the Korean government (MSIT) under Project No. NRF-2008-00458.

\end{document}